\documentclass[prl, twocolumn, 10pt]{revtex4-2}

\usepackage[font=footnotesize, labelfont=bf, justification=centerlast, format=plain, figurename=Figure]{caption}  

\usepackage{xcolor}
\definecolor{darkblue}{rgb}{0.00, 0.00, 0.55}
\definecolor{darkmagenta}{rgb}{0.50, 0.00, 0.50}
\definecolor{darkcerulean}{rgb}{0.03, 0.27, 0.49}

\usepackage{amsmath}
\usepackage{amssymb}
\usepackage{graphicx,epsf}
\usepackage{hyperref}
\hypersetup{
    colorlinks         = true,
    citecolor          = darkblue,
    linkcolor          = darkblue,
    filecolor          = magenta,
    urlcolor           = darkmagenta,
    bookmarks             = false,
    unicode                  = true,
    bookmarksopen       = true,
    bookmarksopenlevel = 1
}

\usepackage[T2A]{fontenc}
\usepackage[cp1251]{inputenc}
\usepackage[english,russian]{babel}
\usepackage{enumitem}
\usepackage{epsfig}
\usepackage{cmap}

\makeatletter
\renewcommand{\fnum@figure}{Figure \thefigure}
\makeatother

\begin{document}
\pagestyle{plain}

\title{Period-dependent suppression of Fourier peaks for topography images:  \\ Analysis of periodicity of triple-step arrays on vicinal Si($h\,h\,m$) surfaces}

\author{A. Yu. Aladyshkin$^{1-3,*}$, A. N. Chaika$^{4,*}$, V. N. Semenov$^{4}$, A. S. Aladyshkina$^{5}$, A. M. Ionov$^{4}$, S. I. Bozhko$^{4}$}

\affiliation{ \\
$^1$Institute for Physics of Microstructures Russian Academy of Sciences, GSP-105, 603950 Nizhny Novgorod, Russia\\
$^2$Center for Advanced Mesoscience and Nanotechnology, Moscow Institute of Physics and Technology, Institutsky 9, 141700 Dolgoprudny, Russia\\
$^3$Lobachevsky State University of Nizhny Novgorod, Gagarin Av. 23, 603022 Nizhny Novgorod, Russia \\
$^4$Osipyan Institute of Solid State Physics Russian Academy of Sciences, Ac. Osipyan str. 2, 142432 Chernogolovka, Russia\\
$^5$National Research University Higher School of Economics (HSE University), 603155 Nizhny Novgorod, Russia \\
$^*$Corresponding authors, e-mail: aladyshkin@ipmras.ru, chaika@issp.ac.ru}

\begin{abstract}
Reliable determination of the periodicity of multiatomic steps on high-Miller-index vicinal surfaces is often complicated by (i) the presence of relatively narrow terraces tilted at large angles relative to the scanning plane, and (ii) unavoidable distortions in the lateral direction resulting from creep and improper calibration of a piezo scanner. This work paves a way to overcome these obstacles using the period-dependent Fourier peaks suppression, which is invariant to affine distortions of the atomically resolved scanning probe microscopy images and to defects of the terrace atomic structure. We argue that the period of the triple-step arrays fabricated on vicinal Si(5\,5\,6) and Si(5\,5\,7) single-crystal wafers can be determined from raw scanning tunneling microscopy data, without preliminary corrections, with precision approaching the interatomic distance. We demonstrate that the intensities of the Fourier peaks in the differential maps, derived from raw topography images and containing stripes with the $7\times7$ reconstruction, strongly depend on the period of the stripes. The suppression of the ninth Fourier peak indicates that the regular array of triple steps formed on the Si(5\,5\,7) wafers has a period of $18b=5.99$\,nm in projection onto the Si(1\,1\,1) terrace plane, where $b=0.333$\,nm is the distance between atomic rows of the $1\times 1$ lattice in the $[\bar{1}\,\bar{1}\,2]$ direction. This means that the nominally (5\,5\,7)-oriented Si wafers may correspond locally to the Si(8\,8\,11) orientation. The method can be applied to the precise analysis of other vicinal surfaces with narrow and wide terraces containing an integer number of Si(1\,1\,1)$7\times7$ unit cells.
\end{abstract}

\maketitle

\section{Introduction}

\begin{figure*}[t!]
\centering
\includegraphics[height=80 mm]{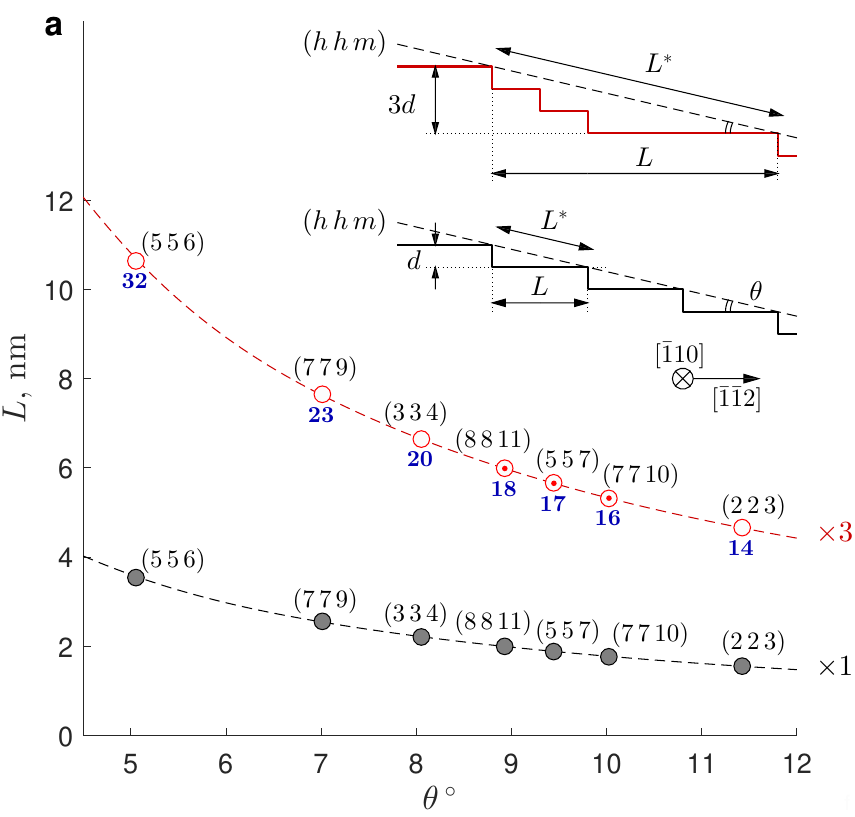}
\includegraphics[height=80 mm]{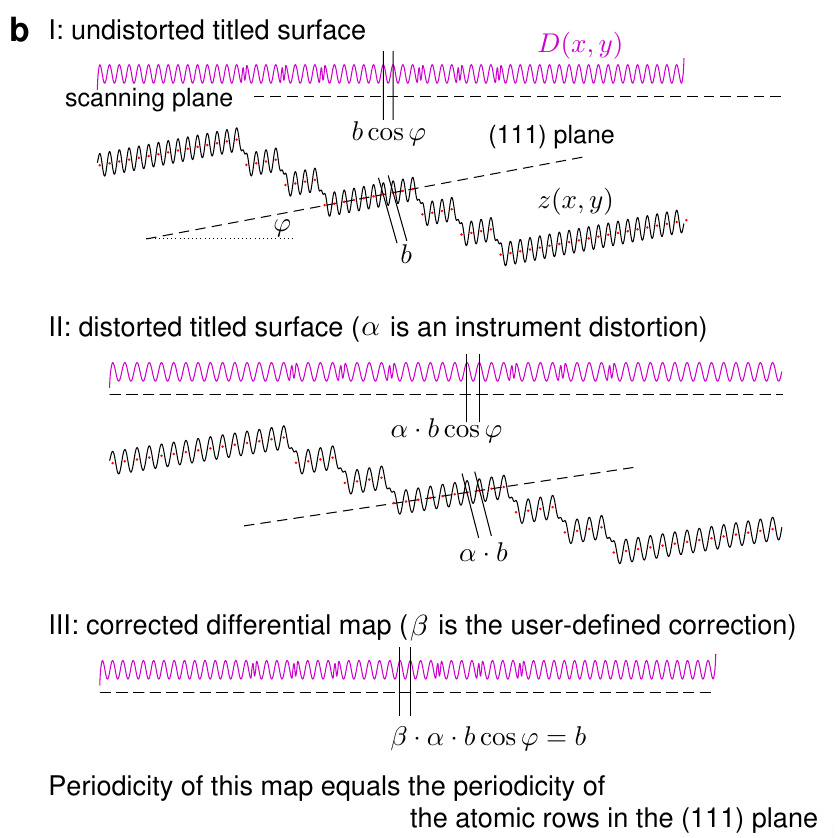}
\caption{{\bf a}~-- Period $L$ of an array consisting of monatomic steps ($\bullet$) and triple steps ($\circ$), aligned along the $[\bar{1}\,1\,0]$ direction, as a function of the miscut angle $\theta$ between the $(h\,h\,m)$ and $(1\,1\,1)$ crystallographic planes for various vicinal Si surfaces \cite{Comment}. Insets illustrate the relationships between the interlayer distance \mbox{$d=c/\sqrt{3}= 0.314\,$nm} and the step spacings: $L = d/\tan\,\theta$ for monatomic steps and \mbox{$L=3d/\tan\,\theta$} for triple steps, where \mbox{$c=0.543$\,nm} is the lattice constant of bulk Si crystals. The introduced step periodicity correspond to those projected onto the (1\,1\,1) plane, and differs from the step periodicity $L^*$ measured along the vicinal surface, where $L^*=L/\cos\,\theta$.  Bold numbers (14, 16, 17, 18, 20, 23, and 32) indicate the triple-step spacings normalized to the distance $b = \sqrt{3}c/(2\sqrt{2})=0.333\,$nm between atomic rows of the $1\times 1$ lattice (see Eq.~(\ref{Eq-ab})). \linebreak {\bf b}~-- Schematic representation of the data processing. The sample surface is typically tilted by an angle $\varphi$ relative to the scanning plane. Case I: a topography image (black line) corresponding to atomic rows (red dots) equally spaced in the (1\,1\,1) plane at a distance $b$ can be converted into a signal with periodic modulation in the difference-of-Gaussians map, which is, in fact, the projection of the atomic features onto the scanning plane (magenta line). The distance between maxima in the differential map is $b \cos\varphi$. Case II: instrumental distortions coherently change both the distance between atoms in the (1\,1\,1) plane and the apparent spacing between atoms in the scanning plane ($\alpha$ is an unknown distortion factor). Case III: by introducing the correction factor $\beta=1/(\alpha \cos\varphi)$ one can obtain the spacing between the maxima in the differential map equal to $b$. The periodicity in the  differential map after correction coincides with that for an ideal (1\,1\,1) plane. This allows us to state that the corrected differential maps can be viewed as a projection of the adatoms at the sample surface onto the (1\,1\,1) plane.
\label{Fig-01}}
\end{figure*}

Vicinal or high-Miller-index surfaces \cite{Oura-03}, which deviate from close-packed low-Miller-index planes by a small angle $\theta$ (figure~\ref{Fig-01}a), can be considered as a regular arrangement of atomically flat terraces separated by mono- or multiatomic steps. Vicinal Si($h\,h\,m$) surfaces show potential interest as semiconductor substrates for fabrication of ordered low-dimensional systems, exhibiting distinct electronic, transport, and magnetic characteristics \cite{Erwin-10,Losio-01,Ahn-PRL-03,Lipton-Duffin-08,Tegenkamp-05,Morikawa-10,Brand-15,Quentin-20,Nita-22}. These terrace-step arrays serve not only as a cost-effective and reproducible quantum standard for measuring distances and heights but also as a promising substrate for depositing ultrathin metallic films and designing quasi-one-dimensional metallic wires \cite{Losio-01,Ahn-PRL-03,Lipton-Duffin-08,Tegenkamp-05,Morikawa-10,Brand-15,Quentin-20}. The step-step interaction can affect the morphology and electronic properties of the fabricated nanostructures \cite{Marchenko}. Since it depends on the distance between steps, precise knowledge of the step periodicity on vicinal substrates with large miscut angles ($\theta\gtrsim 5^{\circ}$) is of particular importance. Among stepped structures fabricated on Si($h\,h\,m$) surfaces, the periodic triple-step staircase on clean Si(5\,5\,7) wafers has attracted significant interest \cite{Kirakosian-APL-01,Henzler-03, Teys-06,Zhachuk-09,Zhachuk-14,Chaika-JAP-09, Chaika-SurfSci-09,Chaika-APL-09, Kim-10,PerezLeon-16,PerezLeon-17,Oh-08,Prevot-12}. It was demonstrated by Kirakosian \emph{et al.} \cite{Kirakosian-APL-01} that Si(5\,5\,7) wafers can be patterned into an atomically precise ordered array of Si(1\,1\,1)$7\times 7$-reconstructed terraces and quantized triple steps (each of height equal to three interplanar spacings). Kirakosian \emph{et al.} \cite{Kirakosian-APL-01} identified the local surface orientation and the triple-step periodicity (i.e., the distance between neighboring Si(1\,1\,1)$7\times7$ terrace edges) as Si(5\,5\,7) and 5.73 nm, measured along the vicinal surface. However, both these findings and the detailed atomic structure of the triple steps have been disputed in several subsequent studies \cite{Henzler-03,Teys-06,Zhachuk-09,Zhachuk-14,Chaika-JAP-09, Chaika-SurfSci-09,Chaika-APL-09, Kim-10,PerezLeon-16,PerezLeon-17,Oh-08,Prevot-12}. Scanning probe microscopy (SPM), X-ray, and low-energy electron diffraction (LEED) studies from different groups have yielded several competing models for the atomic structure of the periodic triple-step array. Moreover, using atomically resolved scanning tunneling microscopy (STM), Teys \emph{et al.} \cite{Teys-06,Zhachuk-09,Zhachuk-14} concluded that the periodicity of the atomically precise triple-step staircase is equal to 5.33 nm in projection to the Si(1\,1\,1) terrace plane (or 5.41\,nm in the vicinal surface plane) and corresponds to the Si(7\,7\,10) rather than the Si(5\,5\,7) surface orientation.

A critical obstacle in the processing of topography images of clean vicinal surfaces arises from the facts that (i) the mean width of atomically flat terraces is rather small (on the order of several nm), and (ii) all terraces are inclined at a large angle ($\sim10^{\circ}$) relative to the scanning plane. These two factors complicate the alignment of topography images using standard plane subtraction, especially in the presence of small-scale atomic corrugation ($\sim 0.01$ nm), uncontrollable drift and creep artifacts, as well as slow variations of background signal. As a result, estimating the plane parameters from a reference area only a few nanometers wide introduces large uncertainties, leading to imperfect compensation of the global sample slope. For example, the typical width of the (1\,1\,1) terraces on the Si(5\,5\,7) \cite{Kirakosian-APL-01} and Si(7\,7\,10) surfaces \cite{Teys-06} corresponds to four rows of adatoms of the Si(1\,1\,1)$7\times7$ reconstruction (or 9 atomic rows of the Si(1\,1\,1)$1\times1$ lattice). A global plane subtraction, frequently used for the analysis of Si($h\,h\,m$) SPM data \cite{Teys-06,Zhachuk-09,Zhachuk-14,Chaika-SurfSci-09,Kim-10,PerezLeon-16,PerezLeon-17,Oh-08}, can provide misaligned topography images with tilted terraces. As a result, such partially aligned images cannot be adequately compared with the parameters of the unit cell of the Si(1\,1\,1)$7\times7$ reconstruction visible on the flat terraces. Numerical differentiation of the topography signal along the fast-scanning direction ($\partial z/\partial x$) for visualizing atomic structure on steps and terraces \cite{Kirakosian-APL-01} can convert raw topography data into a picture with distorted contrast, which does not correspond to the actual positions of the surface atomic features. In order to balance contrast along the multiatomic steps, P\'{e}rez Le\'{o}n \emph{et al.} \cite{PerezLeon-17} have subtracted a linear fit for each line in the raw topography image. However, this method works only if the step edges in the topography image are parallel to the horizontal direction, which is not typical for SPM measurements on stepped surfaces. The aforementioned problems in SPM image processing and the inability to correctly remove drift artifacts may be responsible for disagreements regarding the atomic structures of the triple-step staircases fabricated on nominally (5\,5\,7)-oriented Si wafers. Some authors considered it to be a Si(5\,5\,7) \cite{Kirakosian-APL-01,Oh-08}, Si(7\,7\,10) \cite{Teys-06,Zhachuk-09,Zhachuk-14}, or Si(8\,8\,11) \cite{Chaika-26} staircase, thus estimating a different periodicity (5.65, 5.33, and 5.99 nm in projection onto the Si(1\,1\,1) plane, respectively).
Our work shows a way to solve the problem of precise determination of step periodicity on vicinal surfaces using a straightforward two-stage procedure: first, preparing differential maps from raw topography data by the difference-of-Gaussian approach, and, second, applying Fourier analysis to the differential maps. This procedure is robust to both random atomic defects on the terraces and scaling and affine distortions of the STM images caused by thermal drift, creep of piezoceramics, and improper scanner calibration. Moreover, the results obtained by this procedure become unbiased, since they do not depend on the qualifications of the user who interprets raw data. In particular, the suppression of specific peaks in the Fourier maps obtained from raw topography data allows us to reliably identify that the periodic triple-step staircases fabricated on Si(5\,5\,6) and Si(5\,5\,7) wafers in our experiments correspond to the Si(5\,5\,6) and Si(8\,8\,11) surface orientations, respectively. The proposed method for processing scanning probe microscopy images can be used for precise determination of step periodicity on other vicinal surfaces.

In our recent publications \cite{Aladyshkin-JPCC-24,Aladyshkin-Ultram-24} we demonstrated the effectiveness of the difference-of-Gaussians (DoG) technique in the minimization of tilt and background artifacts and the accentuation of atomic-level details on faceted, stepped, and non-flat surfaces. The idea is simple. The convolution of a noisy two-dimensional topography image $z(x,y)$ with two Gaussian functions of different widths $\sigma^{\,}_1$ and $\sigma^{\,}_2>\sigma^{\,}_1$ is given by
\begin{multline}
\label{Eq-diff-gaussians}
D(x,y) =  \iint  \frac{z(x',y')}{2\pi \sigma^2_1}\,e^{-((x-x')^2+(y-y')^2)/(2\sigma^2_1)}\,dx'dy' \\ - \iint \frac{z(x',y')}{2\pi \sigma^2_2}\,e^{-((x-x')^2+(y-y')^2)/(2\sigma^2_1)}\,dx'dy'.
\end{multline}
This procedure acts as a robust noise-reduction filter, highlighting the local curvature distribution of the slowly varying component of a noisy signal. Since local curvature reaches its maximum at atomic sites, the differential signal $D(x,y)$ provides simultaneous imaging of atomic lattices over the entire surface, independent of local variations in tilt and height (figure~\ref{Fig-01}b). Methods based on the difference between two Gaussian-blurred images are well-established and widely used in digital image processing and computer vision \cite{Gonsales-04,Kovasznayl-53,Marr-80,Young-1987,Lindeberg-15,Assirati-2014} for the automatic detection of edges, interfaces, and defects in grayscale images. As demonstrated in Refs. \cite{Aladyshkin-JPCC-24,Aladyshkin-Ultram-24}, this method corresponds to the projection of the atomic features at the sample surface onto the scanning plane (figure~\ref{Fig-01}b). The approach (\ref{Eq-diff-gaussians}) is highly robust to a particular choice of $\sigma^{\,}_1$ and $\sigma^{\,}_2$. Such filtering allows for local background subtraction and direct visualization of atomic reconstructions on all tilted surface areas for a wide range of $\sigma^{\,}_2/\sigma^{\,}_1$ ratios. Typically, we take $\sigma^{\,}_1$ equal to several pixels ($\sim 0.05-0.1$\,nm) to suppress high-frequency noise, and $\sigma^{\,}_2$ of the order of the interatomic distance ($\sim 0.1-0.2$\,nm) to highlight the spatial variations of the local curvature, which is associated with atomic features both on the terraces and at the step edges.

\begin{figure}[t!]
\centering
\includegraphics[width=85 mm]{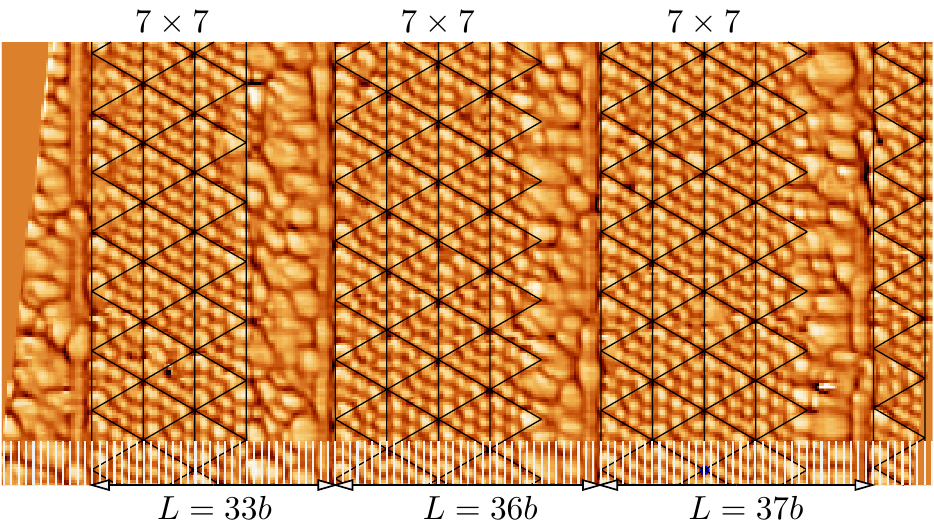}
\caption{A typical map of the differential signal $D(x,y)$ corresponding to the topography image of a vicinal Si(5\,5\,6) surface with rather wide (1\,1\,1) terraces after applying geometrical corrections (corrected image size $42\times 20$ nm$^2$, bias voltage $U = -0.6$ V, tunneling current $I = 80$ pA). This map was prepared using the difference-of-Gaussians procedure (\ref{Eq-diff-gaussians}) as described in the text (smoothing parameters: $\sigma_1 = 0.06$ nm and $\sigma_2 = 2\sigma_1 = 0.11$ nm). This differential map after corrections corresponds to the projection of the surface features onto the (1 1 1) plane (figure~\ref{Fig-01}b). Black solid lines mark the edges of the Si(111)$7\times7$ unit cells. The positions of the atomic rows in the Si(111)$1\times1$ lattice are marked by vertical white lines spaced regularly at $b = 0.333$ nm.
\label{Fig-02}}
\end{figure}

After minimizing tilt and visualizing atomic features on all flat terraces, we face challenges in the accurate determination of the lateral dimensions of the stepped patterns. Figure \ref{Fig-02} shows a typical map of the differential signal $D(x,y)$ obtained from an STM image measured on a Si(5\,5\,6) wafer. To restore orthogonality of the fast and slow scanning directions and correct lateral dimensions, we apply  scale and affine transformations. In order to find the parameters of a proper compensating transformation, we should compare the adatom positions in the corrected image with those in the reference map of the $7\times 7$ reconstruction (figure~\ref{Fig-01}b). When the typical widths of flat terraces with the $7\times 7$ reconstruction are sufficiently large, the comparison can be easily performed with a high precision, yielding a reliable evaluation of the surface structure parameters. In particular, we easily determine the quantized spacings between triple steps in figure~\ref{Fig-02} (11.00, 12.00, and 12.30\,nm) with an accuracy better than 0.05\,nm. However, for the periodic triple-step staircase fabricated on Si(5\,5\,7) wafers with rather narrow flat terraces containing only one $7\times 7$ unit cell, this method of direct comparison of visible atomic features and ones expected for the $7\times 7$ lattice and atomic defects (e.g., missing adatoms) becomes more complicated and demands high accuracy and thorough statistical analysis, taking into account different possible configurations of the triple steps and terraces \cite{Chaika-26}.

In this manuscript, we demonstrate that the differential maps $D(x,y)$ obtained from raw topography STM images by the DoG procedure can be used for accurate determination of the periodicity of the triple-step structures on Si($h\,h\,m$) surfaces, even with narrow flat terraces (less than $2.5$\,nm wide), without any prior correction in either the lateral or vertical direction. The method is based on the analysis of the Fourier patterns of the differential maps and provides ultimate precision in determining the step periodicity. The method is robust to randomly distributed atomic defects, noise, nonzero tilt, and constant drift distortions typical of SPM experiments at room and lower temperatures. The suppression of the peculiar peaks in the Fourier patterns undoubtedly proves that the periodic triple-step staircase reproducibly fabricated on Si(5\,5\,7) wafers corresponds to the Si(8\,8\,11) surface orientation as identified in our recent study \cite{Chaika-26}. We emphasize that the vicinal surfaces Si(8\,8\,11) have not yet been considered in the literature.

\section{Experimental procedure}

Experimental investigations of the crystalline structure of vicinal Si surfaces were carried out in an ultrahigh vacuum (UHV) scanning tunneling microscope \mbox{GPI-300} operating at room temperature and base vacuum pressure of \mbox{$1\cdot 10^{-10}\,$mbar}. All STM experiments were carried out in the constant current mode and constant electrical potential of the sample $U$ with respect to the tip. Electrochemically etched polycrystalline and single-crystalline W tips were employed as STM probes after \emph{in situ} electron bombardment and ion etching \cite{Chaika-SciRep-14,Chaika-EPL-10} in the UHV chamber. STM topography images and DoG maps were processed using the freely available WSXM software \cite{wsxm-ref} and our own code written in Matlab.

Periodic triple-step staircases with different periodicity were fabricated on vicinal samples with typical dimensions of $0.5\times3\times8$\,mm$^3$ cut at different angles from single-crystal Si wafers ($n$-type, P-doped, resistivity of the order of $25 \Omega\cdot$cm at 300 K). These crystals were cut at an angle of either $9.45^{\circ}$ or $5.05^{\circ}$ relative to the (1\,1\,1) plane to obtain Si(5\,5\,7) or Si(5\,5\,6) wafer orientations (figure~\ref{Fig-01}a), respectively. Then these crystals were polished mechanically and chemically \cite{Chaika-Abrosimov}. Although the precision of the wafers orientation was of the order of $0.01^{\circ}$, one cannot exclude minor variations in the local normal direction across the wafer after polishing, and the dynamics of multiatomic steps upon annealing \cite{Latyshev,Yagi}. For example, the nominally (5\,5\,7)-oriented Si wafers may locally correspond to the (8\,8\,11), (5\,5\,7), or (7\,7\,10) crystallographic planes (figure~\ref{Fig-01}a). To minimize the step bunching effects and fabricate periodic, atomically accurate step arrays on large surface areas, we employed in situ direct-current annealing in the UHV chamber, which included the following main steps:

\noindent
(i) continuous outgassing at a moderate temperature of about $600^{\circ}$C for 15--20 hours;

\noindent
(ii) flash heating to $1250-1300^{\circ}$C (several seconds) to remove the protective oxide layer and adsorbates from the surface;

\noindent
(iii) rapid (20--30 seconds) cooling to $900-950^{\circ}$C to avoid step bunching and undesirable growth of Si(1\,1\,1) terraces at intermediate temperatures \cite{Yagi};

\noindent
(iv) annealing for approximately 1 minute at $900-950^{\circ}$C to stabilize a uniform step distribution across the surface;

\noindent
(v) gradual (15–20 minutes) cooling from $900^{\circ}$C to $600-650^{\circ}$C to form the $7\times7$ reconstruction on the terraces and to convert monatomic steps to triple steps;

\noindent
(vi) continuous (from 20 minutes to 2 hours) post-annealing at $600-650^{\circ}$C.

The current was directed perpendicular to the step edges in the step-up direction, as this provided better reproducibility and higher regularity of the triple-step arrays on Si($h\,h\,m$) wafers in our experiments. Using the above described procedure we could reproducibly fabricate periodic triple-step arrays on relatively large surface areas.

We prepared several dozen samples on Si(5\,5\,7) and Si(5\,5\,6) wafers; the structural properties of these samples have been studied in detail over the last five years \cite{Chaika-26}. As a result, we acquired hundreds of topography images under various conditions. For all processed images, the periodicity of the triple-step arrays was determined manually by a standard procedure involving compensated transformations (similar to the one shown in figure \ref{Fig-02}). To verify the theoretical predictions formulated below, we selected two topography images, one for the Si(5\,5\,7) sample (figure \ref{Fig-07}) and one for the Si(5\,5\,6) sample (figure \ref{Fig-10}). The criteria for this particular selection were that (i) the distances between similar steps in the area of interest should be equal, and (ii) the flat (1\,1\,1) terraces should contain one and three complete unit cells of the $7\times7$ reconstruction, respectively, without additional rows of adatoms \cite{Chaika-26}. To show the variety of triple-step arrangements on the same sample, we present a topography image, displaying different distances between similar steps (figure \ref{Fig-08}).

\section{Results and discussion}

\subsection{Model: periodic array of $7\times 7$ stripes}

\begin{figure}[h!]
\begin{center}
\includegraphics[width=85 mm]{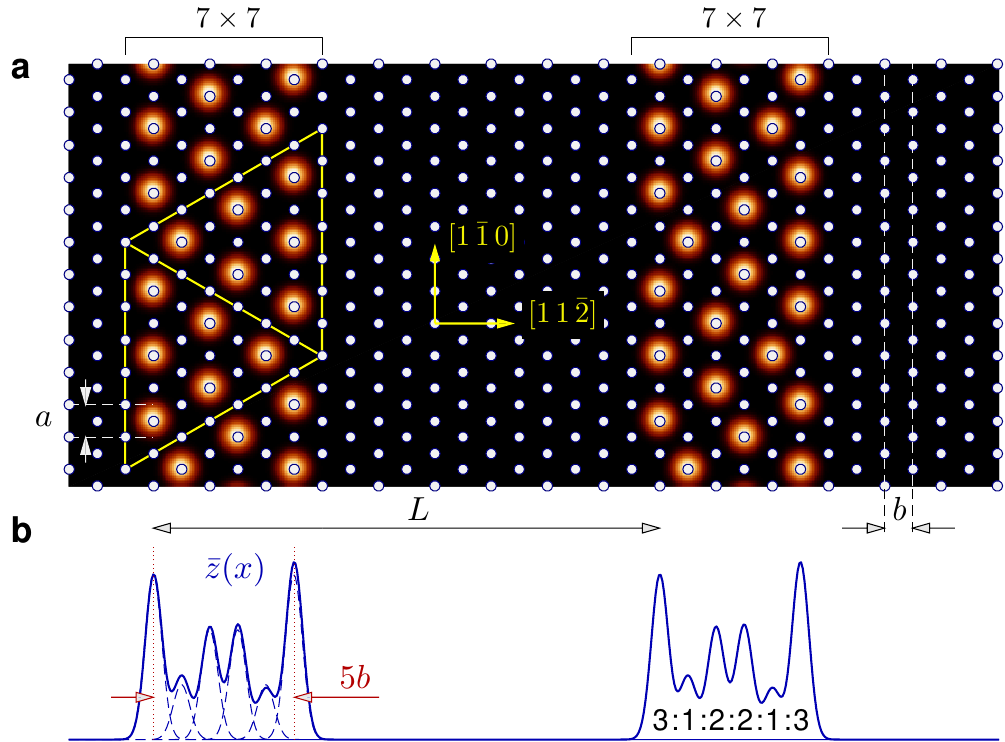}
\end{center}
\caption{\textbf{a}~-- Periodic model structure which consists of the Si(1\,1\,1)$7\times 7$ stripes, repeating with a period $L$ (here $L = 18 b$). Yellow solid lines outline a unit cell for the $7\times 7$ lattice, consisting of two equilateral triangles and containing twelve adatoms. White dots represent the positions of the Si(1\,1\,1)$1\times 1$ atoms. \textbf{b}~-- The local concentration of the adatoms, averaged along the vertical direction, is plotted as a function of the horizontal $x$-coordinate.
\label{Fig-03}}
\end{figure}

We begin by considering a simple model problem. We introduce one-dimensional periodic arrays consisting of flat terraces with the Si(1\,1\,1)$7 \times 7$ reconstruction.  Such $7\times 7$ stripes are oriented along the vertical $[1\,\bar{1}\,0]$ direction and evenly spaced along the horizontal $[1\,1\,\bar{2}]$ direction with a period $L$ (figure~\ref{Fig-03}a). One can calculate the distances between neighboring atoms within $[1\,\bar{1}\,0]$-oriented atomic rows of the $1\times 1$ lattice and between these rows
\begin{eqnarray}
\label{Eq-ab}
a = \frac{c}{\sqrt{2}} = 0.384\,{\rm nm} \quad \mbox{and} \quad b = \frac{\sqrt{3}}{2} a = 0.333\,{\rm nm},
\end{eqnarray}
where $c=0.543\,$nm is the lattice constant for bulk Si crystal. We limit our analysis to commensurable arrays where the positions of the adatoms for the $7\times 7$ reconstruction coincide with those of the $1\times 1$ lattice. Specifically, as figure~\ref{Fig-03}a illustrates, we consider stripes composed of one complete rhombic unit cell of the $7\times 7$ lattice. The distance between the centers of the left and right rows of adatoms equals $5b=1.66\,$nm.

\begin{figure*}[t!]
\begin{center}
\includegraphics[width=43 mm]{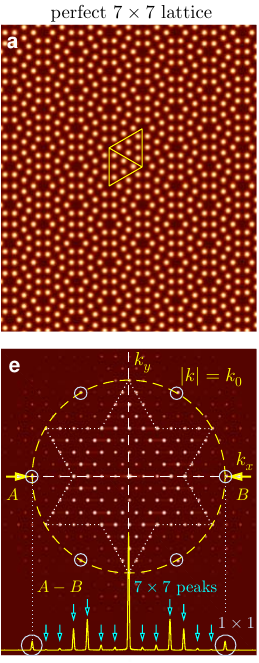}
\includegraphics[width=43 mm]{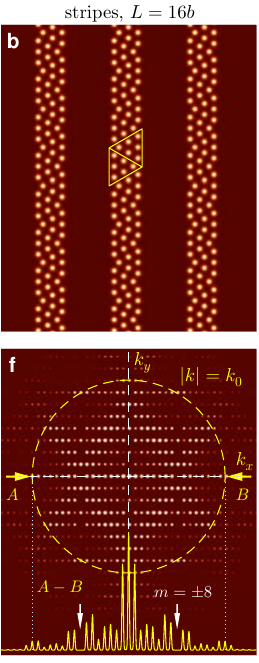}
\includegraphics[width=43 mm]{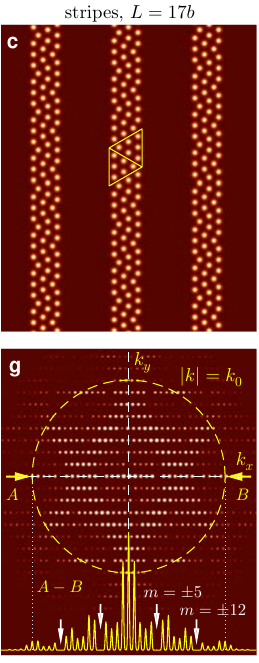}
\includegraphics[width=43 mm]{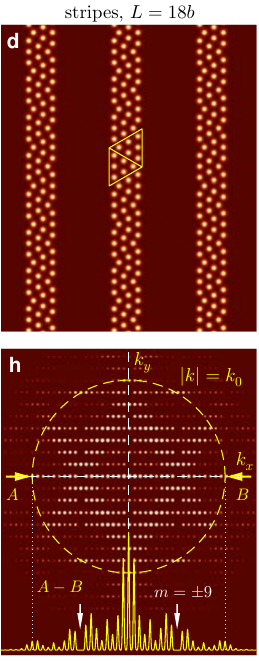}
\end{center}
\caption{\textbf{Top row}~-- Model images $z(x,y)$ for the perfect Si(1\,1\,1)$7\times 7$ lattice (panel a) and for one-dimensional periodic arrays of stripes with the Si(1\,1\,1)$7\times 7$ reconstruction and periods $L=16 b$ (panel b), $17 b$ (panel c), and $18 b$ (panel d). These parameters correspond to the atomic lattices of the Si(1\,1\,1) terraces for Si(7\,7\,10), Si(5\,5\,7), and Si(8\,8\,11) wafers, respectively, after projection onto the (1\,1\,1) plane. All images have dimensions $18\times 21.6$\,nm$^2$. The parameter $b=0.333\,$nm is the spacing between atomic rows in the horizontal $[1 1 \bar{2}]$ direction. \linebreak
\textbf{Bottom row}~-- Fourier transforms $|\hat{z}(k^{\,}_x,k^{\,}_y)|$ computed for the images in panels a--d (image size $50\times 60$\,nm$^{-2}$). To perform Fourier analysis correctly, we prepare model images of size $100\times 100$\,nm$^2$, apply a Gaussian window to them to eliminate finite-size effects, and then calculate the Fourier transforms. The radius of the circle marking the expected positions of the first-order Fourier peaks for the Si(1\,1\,1)$1\times 1$ lattice, is equal to $k^{\,}_0=18.89\,$nm$^{-1}$ [see Eq.~(\ref{Eq-k0})]. Superstructure reflexes for the $7\times 7$ reconstruction are indicated by green arrows (panel e). Insets show cross-sections along the $A-B$ line, and white vertical arrows indicate the fully suppressed Fourier peaks.
\label{Fig-04}}
\end{figure*}

Figure~\ref{Fig-04} shows a reference image displaying regular Si(1\,1\,1)$7\times 7$ reconstruction (panel a) and its Fourier transform (panel e). It is easy to observe six intense Fourier peaks marked by open dots, arranged along a circle with a radius of
\begin{eqnarray}
\label{Eq-k0}
k^{\,}_0 = \frac{2\pi}{b} = \frac{4\pi}{\sqrt{3} a}= 18.89\,\mbox{nm}^{-1}.
\end{eqnarray}
These peaks clearly correspond to the first-order Fourier peaks for the hexagonal Si(1\,1\,1)$1\times 1$ lattice. The presence of the $7\times 7$ lattice results in the formation of satellite peaks marked by arrows (panel e). The distance between these satellite $7\times 7$ peaks along the horizontal axis equals $k^{\,}_0/7$.

\begin{figure*}[t!]
\begin{center}
\includegraphics[width=43 mm]{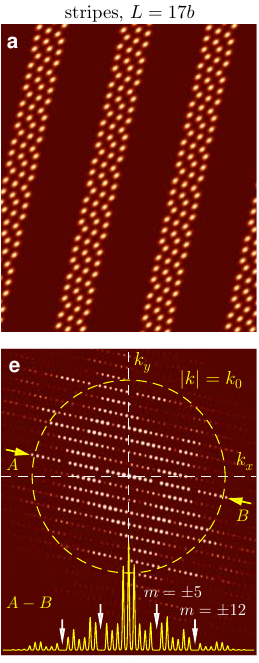}
\includegraphics[width=43 mm]{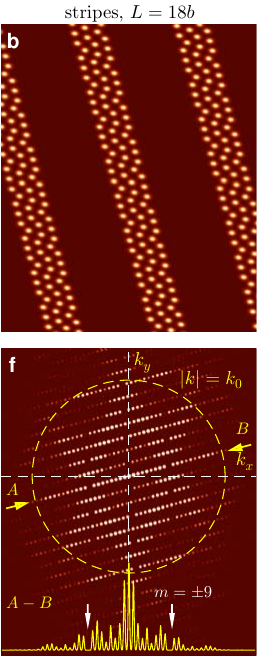}
\includegraphics[width=43 mm]{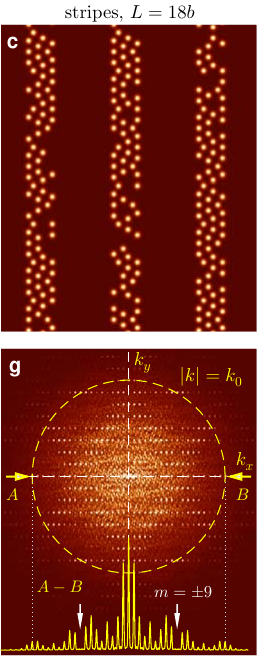}
\includegraphics[width=43 mm]{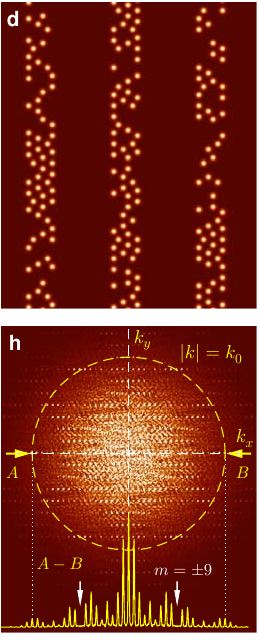}
\end{center}
\caption{\textbf{Top row}~-- Model images $z(x,y)$ displaying one-dimensional periodic arrays of the $7\times 7$ stripes with periods $L=17 b$ (panel a) and $18 b$ (panel b) after applying the scaling and affine transformations, and of undistorted periodic arrays of the $7\times7$ stripes with period 18b containing different fractions of randomly distributed vacancy-type defects: 20\% (panel c) and 40\% (panel d). \textbf{Bottom row}~-- Fourier transforms $|\hat{z}(k^{\,}_x,k^{\,}_y)|$ computed for the images in panels a-d. All dimensions and notations are identical to those in figure~\ref{Fig-04}. The presence of random defects increases the intensity of the non-resonant background without visible suppressing the amplitudes of the resonant peaks.
\label{Fig-05}}
\end{figure*}

We proceed with the analysis of periodic arrays of the $7\times 7$ stripes with periods $L =16b$ (panel b),  $17b$ (panel c), and $18b$ (panel d). Such arrays can correspond to fragments of the Si(1\,1\,1)$7\times 7$ lattice revealed by the DoG procedure, on atomically flat terraces of of the Si(5\,5\,7),  Si(7\,7\,10), and Si(8\,8\,11) triple step staircases, respectively. Fourier transforms of these model images are shown in panels f--h. The structure of the $k$-space of the periodic array of the $7\times 7$ stripes becomes rather complicated: additional Fourier maxima emerge, with their positions determined by the period $L$. One can anticipate interference effects due to commensurability between different periods in the $k$-space ($k^{\,}_0/7$ vs. $k^{\,}_0/L$). A crucial finding is the period-dependent suppression of the specific Fourier peaks positioned at $k^{\,}_y=0$. Particularly, we predict a complete suppression of the $m$-th order Fourier peaks in the following cases:
\begin{itemize}
  \item $m=\pm 8$ when $L=16 b$ (figure \ref{Fig-04}f);
  \item $m=\pm 5$ and $\pm 12$ when $L=17 b$ (figure \ref{Fig-04}g);
  \item $m=\pm 9$ when $L=18 b$ (figure \ref{Fig-04}h).
\end{itemize}
It is important to note that this observation remains valid even for images distorted by scaling or affine transformations, which alter the visible dimensions and introduce non-orthogonality between the fast and slow scanning directions (figure \ref{Fig-05}a,b), as well as for structures with missing adatoms (figure \ref{Fig-05}c,d). Scaling or affine transformations simulate typical distortions in scanning probe microscopy measurements due to an improperly calibrated piezo-scanner and constant drift. The latter factor is especially important in measurements at room temperature. Thus, FFT-based analysis of the DoG patterns allows us to easily determine the periodicity of the $7\times 7$ stripes by examining the sequential number(s) of the suppressed Fourier peaks even without requiring prior time-consuming correction of geometric distortions in the analyzed topography images \cite{Chaika-26}. For example, a similar effect -- the suppression of the fourth-order Fourier peaks in the X-ray diffraction patterns of DNA molecules -- made it possible to determine the relative shift between the two helices \cite{Franklin}.

To illustrate the origin of this interference effect, we consider a simplified one-dimensional problem. We assume that the density of adatoms averaged along atomic rows (figure~\ref{Fig-03}b) as a function of the $x$-coordinate along the $[1\,1\,\bar{2}]$ direction can be written in the following form
\begin{multline}
\label{Eq-aver-z}
\hspace*{-4mm} \bar{z}(x) = \sum_m \left\{3  e^{-(x+5b/2-mL)^2/(2\sigma^2)}  + e^{-(x+3b/2-mL)^2/(2\sigma^2)}\right.  \\
\hspace*{-14mm} + 2 e^{-(x+b/2-mL)^2/(2\sigma^2)} + 2 e^{-(x-b/2-mL)^2/(2\sigma^2)} \\  + \left. e^{-(x-3b/2-mL)^2/(2\sigma^2)} + 3 e^{-(x-5b/2-mL)^2/(2\sigma^2)} \right\},
\end{multline}
where $m$ is the summation index. The parameter $\sigma$ in Eq.~(\ref{Eq-aver-z}) represents the width of individual maxima in the spatial distribution of the averaged adatom concentration. The relative amplitudes of the components in Eq.~(\ref{Eq-aver-z}) agree with the $3:1:2:2:1:3$ rule (figure~\ref{Fig-03}b). This rule remains valid even for the $7\times 7$ stripes with defects provided that the missing adatoms are uniformly distributed over the x,y-plane (figure \ref{Fig-05}c,d).

\begin{figure}[t!]
\begin{center}
\includegraphics[width=85 mm]{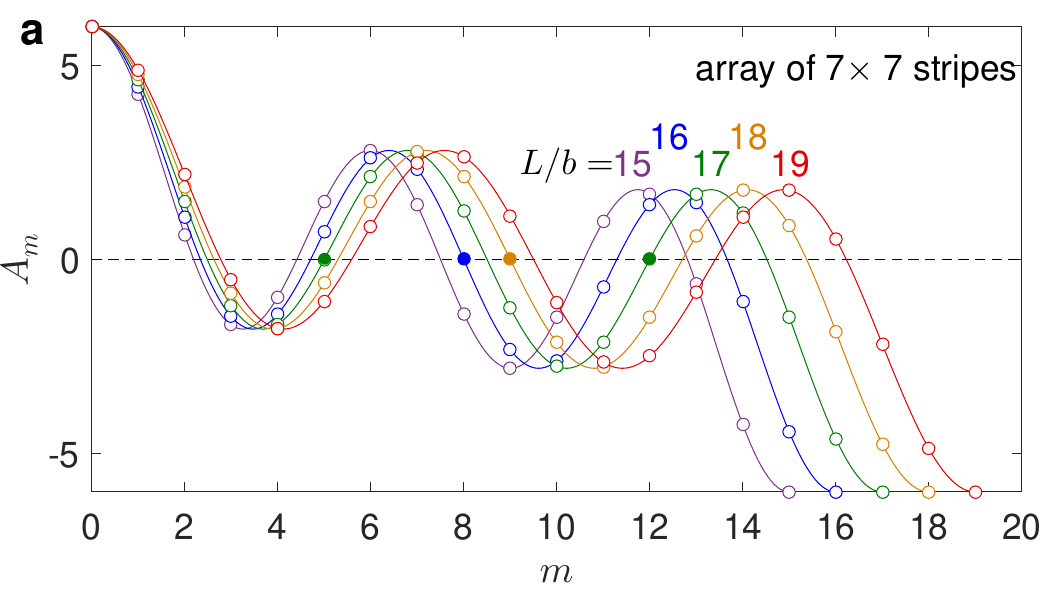}
\includegraphics[width=85 mm]{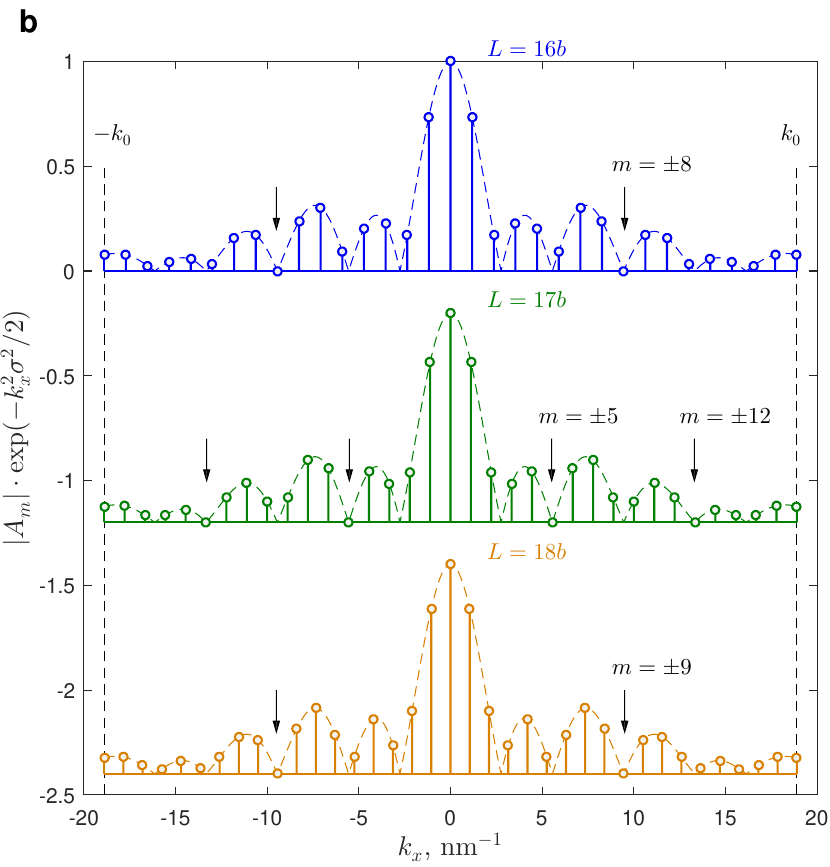}
\end{center}
\caption{\textbf{a}~-- The dependence of the amplitude $A^{\,}_m$ of the $m$-th Fourier maximum on its index for various periods $L$ is represented by open circles according to Eq.~(\ref{Eq-Amplitudes}). The zeros of $A^{\,}_m$ are highlighted by filled circles of the corresponding colors. \textbf{b}~-- The dependence of the normalized amplitude of the $m$-th Fourier peak on the $m$-th value of the wave vector, calculated for arrays of various periods $L$. These charts can be compared with those plotted in figure \ref{Fig-04} (panels f-h).
\label{Fig-06}}
\end{figure}

\begin{figure*}
\begin{center}
\includegraphics[width=170 mm]{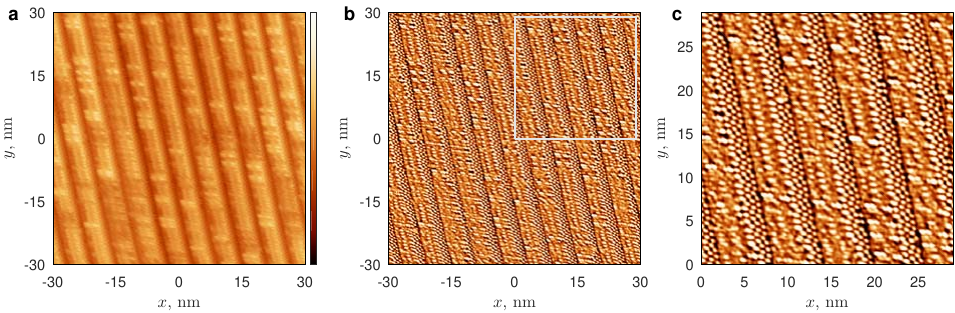}
\includegraphics[width=170 mm]{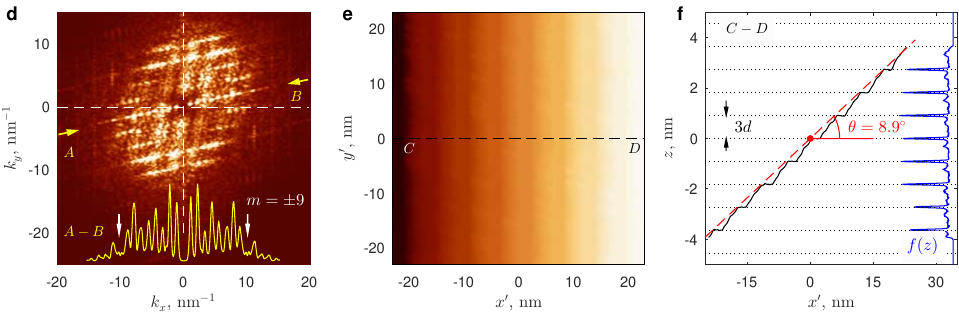}
\end{center}
\caption{\textbf{a}~-- Topography image $z(x,y)$ of the triple step staircase fabricated on vicinal Si(5\,5\,7) wafer after global plane subtraction (nominal image size $61.4\times 61.4\,$nm$^2$, bias voltage $U = 1.5\,$V, tunneling current $I = 60\,$pA).  \textbf{b, c}~-- Maps of the differential signal $D(x,y)$, visualizing the fragments of the $7\times 7$ reconstruction at atomically-flat Si(1\,1\,1) terraces. These images are obtained from the raw image in panel a using the difference-of-Gaussians procedure (\ref{Eq-diff-gaussians}) without additional image processing; filter parameters are equal to $\sigma^{\,}_1=0.1$\,nm and $\sigma^{\,}_2 =  1.7 \sigma^{\,}_1=0.18\,$nm. \textbf{d}~-- Fourier transform $|D(k^{\,}_x, k^{\,}_y)|$ corresponding to the differential signal in the right part of panel b, where $7\times 7$ stripes exhibit greater uniformity. Inset shows the cross-section of the Fourier transforms along the $A-B$ line. \textbf{e}~--  Aligned topography image for the image in panel a after plane subtraction and 12$^{\circ}$ clockwise rotation. \textbf{f} -- Cross-sectional view along $A-B$ line (perpendicular to the steps) combining with normalized probability density functions $f(z)$ (depicted as a blue solid line), characterizing the height distributions for the aligned image.
\label{Fig-07}}
\end{figure*}

\begin{figure}[th!]
\begin{center}
\includegraphics[width=90 mm]{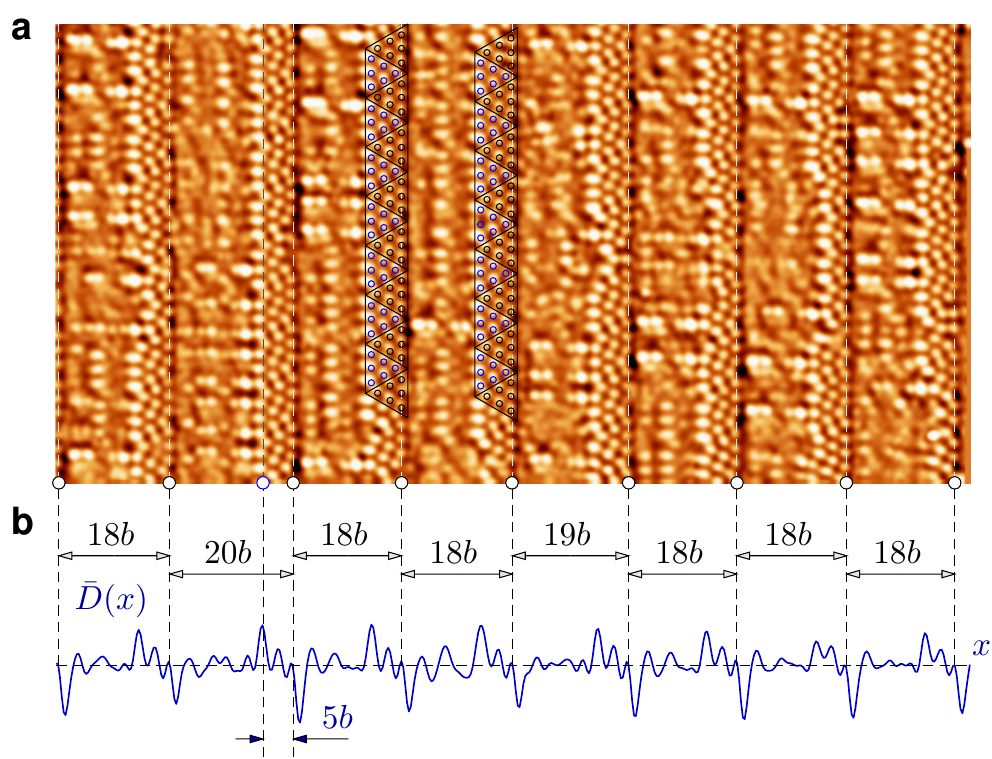}
\end{center}
\caption{\textbf{a}~-- The map of the differential signal $D(x,y)$ prepared using the DoG procedure (\ref{Eq-diff-gaussians}), corresponds to the central region of the image shown in figure \ref{Fig-07}e after applying geometrical corrections (corrected image size $50\times 25\,$nm$^2$, bias voltage $U = 1.5\,$V, tunneling current $I = 60\,$pA). Blue and black dots represent the expected positions of the adatoms for the Si(1\,1\,1)$7\times 7$ lattice within one of the atomically flat terraces. \textbf{b}~-- The differential signal $\bar{D}(x)$, averaged along the vertical direction. Using the positions of the atomic rows at the right edge of each terrace as reference points, we can estimate the distance between neighboring terraces ($b\simeq 0.333$\,nm is the mean distance between atomic rows).
\label{Fig-08}}
\end{figure}

\begin{figure}[th!]
\begin{center}
\includegraphics[width=80 mm]{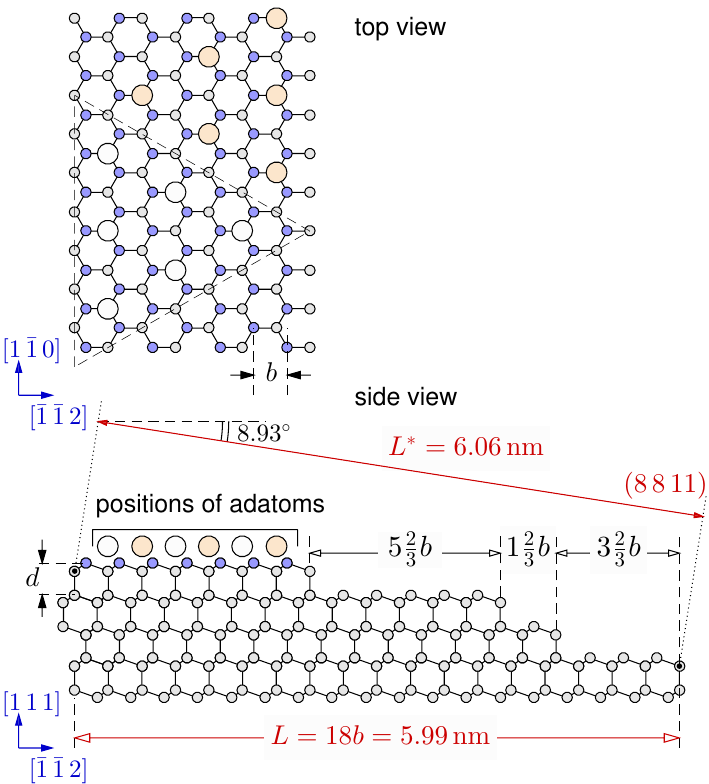}
\end{center}
\caption{One of potential configurations corresponding the stepped Si(8\,8\,11) surface with the period $L=18b$. Two grey atoms marked by black dots are equivalent atoms for the periodic triple-step staircase.
\label{Fig-09}}
\end{figure}

Since the Fourier transform of the Gaussian function in the $x$-domain is the Gaussian function in the $k^{\,}_x$-domain:
\begin{eqnarray}
\nonumber
\frac{1}{\sqrt{2\pi}} \int\limits_{-\infty}^{\infty} e^{-(x'-x_0)^2/(2\sigma^2)} e^{-ik_xx'} dx' = \sigma e^{- ik_xx_0-k_x^2\sigma^2/2},
\end{eqnarray}
we come to the expression for the Fourier transform of the averaged function $\bar{z}(x)$
\begin{multline}
\nonumber
\mathcal{F}[\bar{z}(x)] = \hat{\bar{z}}(k^{\,}_x)  = 2\sigma e^{-k^2_x\sigma^2/2} \sum_me^{-ik^{\,}_x mL} \times  \\  \left\{3\cos \left(\frac{5}{2}k^{\,}_xb\right) + \cos \left(\frac{3}{2}k^{\,}_xb\right)  +  2\cos \left(\frac{1}{2}k^{\,}_xb\right) \right\}.
\end{multline}
Taking into account the identity
\begin{eqnarray}
\nonumber
\sum_{m}  e^{-i k_x mL} = \frac{2\pi}{L} \sum_{m} \delta\left(k^{\,}_x - \frac{2\pi m}{L}\right),
\end{eqnarray}
we arrive at the analytical expression
\begin{eqnarray}
\label{Eq-aver-FT3}
\hat{\bar{z}}(k^{\,}_x) = \frac{4\pi\sigma}{L} e^{-k^2_x\sigma^2/2} \sum_{m} A^{\,}_m \delta\left(k^{\,}_x - \frac{2\pi m}{L}\right),
\end{eqnarray}
where the relative amplitude of the $m$-th Fourier peak equals
\begin{multline}
\label{Eq-Amplitudes}
A^{\,}_m = 3\cos \left(\frac{5\pi m b}{L}\right) + \cos \left(\frac{3\pi m b}{L}\right)  + \\ + 2\cos \left(\frac{\pi m b}{L}\right).
\end{multline}
Figure \ref{Fig-06} shows the dependence of $A^{\,}_m$ on its index $m$ for various periods from $L=15 b$ to $L=19 b$.

Equation (\ref{Eq-Amplitudes}) allows us to draw the following conclusion:

(i) the periodic arrays of $5b$-wide $7\times 7$ stripes with an even $L/b$ ratio always exhibit complete suppression of the Fourier peaks with the indices $m=\pm L/(2b)$, located at $k^{\,}_x=\pm k^{\,}_0/2$. This explains the disappearance of the peaks corresponding to $m=\pm 8$ for the array with $L=16 b$ and $m=\pm 9$ for the array with $L=18b$ 
[see panels f, h in figure~\ref{Fig-04}, and panel f in figure~\ref{Fig-05}].

(ii) for periodic arrays of $5b$-wide $7\times 7$ stripes with an odd $L/b$ ratio, the interference-induced suppression of the Fourier peaks strongly depends on the period. One can anticipate the disappearance of the peaks corresponding to $m=\pm 5$ and $\pm 12$ for the array with $L=17 b$, which agrees with the results of the two-dimensional FFT analysis [see figure~\ref{Fig-04}g and figure \ref{Fig-05}e]. In contrast, complete suppression of the Fourier peaks for the arrays with $L=15b$ and $19b$ appears unlikely.

Thus, this remarkable period-dependent suppression of the Fourier maxima at $k^{\,}_y = 0$ is a direct consequence of the specific relationship among the amplitudes of the maxima in the averaged atomic concentration. Adding (or removing) atomic rows to specific patterns of the $7\times 7$ stripes would clearly alter the profile of the averaged adatom concentration, thereby modifying the effect of the disappearance of specific Fourier peaks. The effect of various periodic patterns composed of $5\times 5$, $7\times 7$, and $9\times 9$ reconstructions can be easily analyzed using the same methodology.

\subsection{Experiment: nominal Si(5\,5\,7) surface becomes a Si(8\,8\,11) surface in reality}

Figure~\ref{Fig-07}a displays one of the typical topography images $z(x,y)$ acquired on the periodic triple step staircase fabricated on the vicinal Si(5\,5\,7) wafer. The map of the differential signal $D(x,y)$ allows us to visualize fragments of the $7\times 7$ reconstruction across all terraces, regardless of their absolute height or local slope variations (figure~\ref{Fig-07}b, c). It should be emphasized that the $7\times 7$ stripes in the right part of panel b display enhanced uniformity and have widths of $5b=1.66\,$nm on average (as seen in panel c). This is fully consistent with the $5b$-wide $7\times 7$ patterns considered in the preceding theoretical section.

The fast Fourier transform applied to the right part of the differential signal in panel b is shown in figure~\ref{Fig-07}d. Recognizing any of the $1\times 1$ peaks in the $k$-domain would successfully resolve the calibration challenge. However, reliable identification of these peaks within the considered dataset is hindered by noise. That is why we propose to use the effect of interference-induced suppression of the Fourier peaks for  calibration purposes. Indeed, the ninth Fourier peak is strongly suppressed, whereas the fifth, eighth and tenth Fourier peaks remain notably intense. The nonzero intensity of the Fourier transform at the position of the ninth peak can be attributed to the presence of local defects like missing adatoms at the terraces and atoms between steps. By comparing these findings with our theoretical results (figures \ref{Fig-04} and \ref{Fig-05}), we readily deduce two key points: (i) the period $L$ of the superstructure in this area is close to $18b$ or 5.99\,nm; (ii) the local surface orientation corresponds to the (8\,8\,11) crystallographic plane. These conclusions are not consistent with the Si(5\,5\,7) \cite{Kirakosian-APL-01,Oh-08} and Si(7\,7\,10) structures \cite{Teys-06,Zhachuk-09,Zhachuk-14} discussed earlier in the literature. We would like to emphasize that our model considers an ordered array of stripes containing an integer number of $7\times 7$ unit cells. For real systems exhibiting different periods and well-ordered linear defects, such as rows of adatoms on the terraces and between the stripes, unambiguous identification of periodicity by this simple Fourier-based analysis seems impossible without additional procedures.

Using this $L=5.99$\,nm value as a reference value, we can rescale the raw image and effectively minimize all geometrical distortions in the lateral plane. After subtraction of a plane from the distortion-free image we can generate an aligned topography image revealing a series of parallel multiatomic steps (figure~\ref{Fig-07}e). A cross-section of the aligned image along the $C-D$ line, shown in figure~\ref{Fig-07}e, exhibits a step-like profile. The histogram (\emph{i. e.} the probability density function $f(z)$ characterizing the visible heights for the aligned image), as anticipated, has a series of narrow maxima whose widths \mbox{($\sim 0.1\,$nm)} are significantly smaller than the intervals between adjacent maxima \mbox{($\sim 0.94\,$nm)}. Consequently, the majority of these steps are triple steps because their heights closely match the theoretical value $3d\simeq 0.941$\,nm, where $d=0.314\,$nm is the interplanar spacing for the Si(1\,1\,1) surface. Clearly the slope of the aligned surface approaches 8.9$^{\circ}$, which corresponds well to the miscut angle of the (8\,8\,11) plane (figure \ref{Fig-01}).

Figure \ref{Fig-08} displays the map of the differential signal corresponding to the topography image in figure~\ref{Fig-07} after rotation, scaling, and affine transformations that restore accurate dimensions in the lateral plane. To demonstrate the correctness of all transformations, we present the anticipated positions of the adatoms for the Si(1\,1\,1)$7\times 7$ lattice (blue and black dots in panel a) for an atomically flat terrace. The differential signal $\bar{D}(x)$, averaged along the vertical direction as a function of the transverse $x$ coordinate, is presented in figure~\ref{Fig-08}b. By measuring the interval between the maxima of $\bar{D}(x)$ within the flat terrace and associating this value with five inter-row spacings $b$, we can determine the periodicity of the neighboring terraces with the $7\times 7$ reconstruction: $18b=5.99\,$nm,  $19b=6.32\,$nm, and $20b=6.65\,$nm. It is clear that the majority of triple steps exhibit a periodicity of 5.99\,nm, consistent with the results obtained from the FFT-based analysis of uncorrected images. In addition, this finding is confirmed by detailed analysis of the differential maps after normalization to the Si(1\,1\,1)$7\times7$ unit cell for a variety of triple step structures observed on this periodic triple step staircase \cite{Chaika-26}. A possible model of triple-step configuration for the Si(8\,8\,11) surface with the period $18b$ is plotted in figure~\ref{Fig-09}.

Even though the triple-step array with a preferential interstep distance of $L = 18b$, fabricated on nominally (5\,5\,7)-oriented Si wafers, spanned micron-sized surface areas \cite{Chaika-26}, we cannot completely exclude breakdowns in the triple-step periodicity. Such breakdowns may correspond to interstep distances that deviate by one or, very rarely, two interatomic distances (e.g., $19b$ and $20b$ in figure 8), and may involve the formation of multistep bunches \cite{Latyshev} in some surface areas needed to ensure the global miscut angle of $9.45^{\circ}$. All these deviations from precise periodicity indicate that the energetics of vicinal surfaces is nontrivial, and different surface phases can be metastable and coexist. This problem is beyond the scope of this paper and will be addressed elsewhere.

\begin{figure*}[t!]
\centering
\includegraphics[width=150 mm]{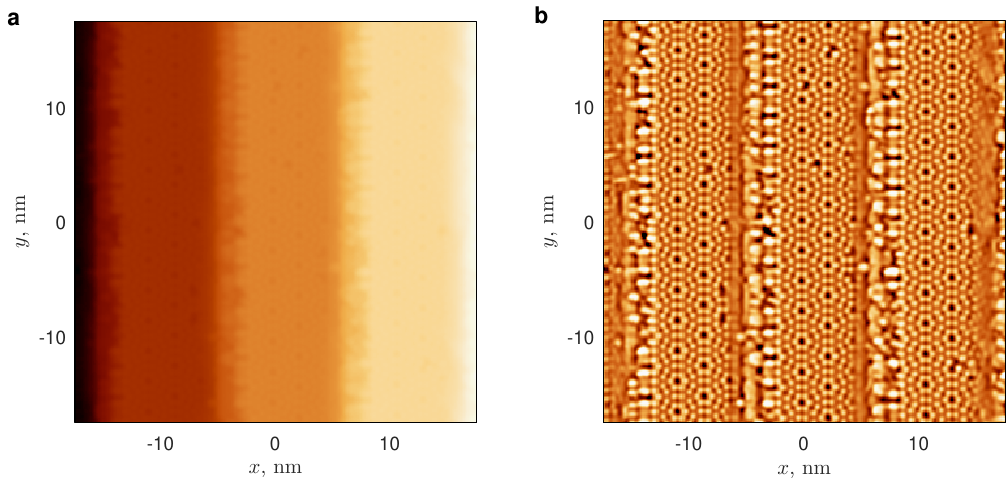}
\includegraphics[width=150 mm]{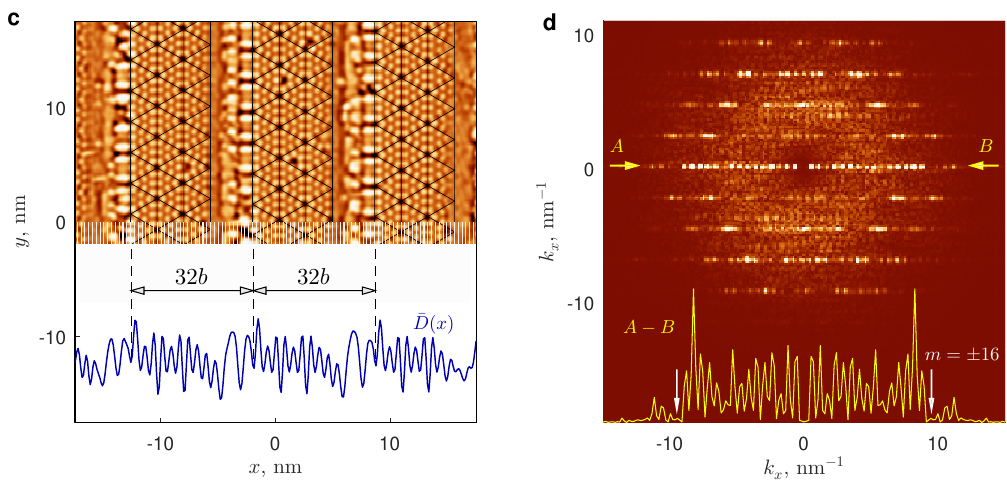}
\caption{\textbf{a, b}~-- Aligned topography image $z(x,y)$  (panel a) and the map of the differential signal $D(x,y)$ (panel b) for the Si(5\,5\,6) surface (corrected image size $35\times 35\,$nm$^2$, bias voltage $U = +1.2\,$V, tunneling current $I = 50\,$pA). Filter parameters for preparing the DoG map (panel b) are equal to $\sigma^{\,}_1=0.12$\,nm and $\sigma^{\,}_2 =  1.5 \sigma^{\,}_1 = 0.18$\,nm. This differential map after corrections corresponds to the projection of the surface features onto the (1\,1\,1) plane (figure~\ref{Fig-01}b).
\textbf{c}~-- Fragment of the map of the differential signal and the array of lines depicting the edges of the Si(1\,1\,1)$7\times 7$ unit cells. The positions of the atomic rows in the Si(111)$1 \times 1$ lattice are marked by vertical white lines spaced regularly at $b=0.333$\,nm [see Eq.~(\ref{Eq-ab})]. The blue line shows the differential signal $\bar{D}(x)$, averaged along the vertical direction.
\textbf{d}~-- Fourier transform $|D(k^{\,}_x, k^{\,}_y)|$ corresponding to the differential signal in the central part of panel b. The inset shows the cross-section of the Fourier transform along the $A-B$ line and highlights the absence of the 16th peak at $k^{\,}_x=\pm 9.44\,$nm$^{-1}$ and $k^{\,}_y=0$.
\label{Fig-10}}
\end{figure*}

A possible model of triple-step configuration for the Si(8\,8\,11) surface with the period $18b$ is plotted in figure~\ref{Fig-09}. Other types of triple-step structures of the same periodicity is presented in Ref.~\cite{Chaika-26}.

\subsection{Experiment: Si(5\,5\,6) surface}

Obviously, this equation (\ref{Eq-Amplitudes}) for the relative amplitudes of the Fourier peak intensities will be the same if the model periodic structure contains $7\times 7$ stripes with an integer number of unit cells (e.g., 2, 3, 4, etc.). An example of such a structure is shown in figure~\ref{Fig-10}, where an aligned $z(x,y)$ STM image (panel a) and the differential map (panel b) of the triple step staircase fabricated on a Si(5\,5\,6) wafer are presented. This surface area contains three consecutive sequences of the triple steps and Si(1\,1\,1) terraces with widths corresponding to three $7\times 7$ unit cells. According to the averaged cross-section of the fragment of the differential map (panel c), the periodicity of the triple-step array in the projection onto (1\,1\,1) plane equals $32b$ or 10.66 nm. According to Eq.~(\ref{Eq-Amplitudes}), a periodic array of wide Si(1\,1\,1)$7\times 7$ stripes with this periodicity indeed provides a Fourier pattern with suppressed peaks with $m=\pm 16$ at $k^{\,}_y=0$ (panel d). This example illustrates that the proposed method for the determination of the periodicity of multiatomic steps based on the analysis of the Fourier peak intensities can be applied to a broad class of stepped surfaces, not only for Si(5\,5\,7) and Si(8\,8\,11).

We can easily obtain the same period by considering the differential map after affine compensation transformation and comparing the positions of the adatoms with those of the ideal Si(1\,1\,1)$7\times 7$ lattice. For a triple-step staircase on Si(5\,5\,6) wafers the difference between adjacent possible periods (e.g., between $32 b$ and $33 b$ or between $32 b$ and $31 b$) corresponds to a relative error of about 3 percent, which requires almost perfect matching of the experimental features with the ideal $7\times 7$ lattice during the image correction procedure. A comparable or even larger uncertainty (on the order of several percent) in the determination of lateral dimensions of the studied stepped atomic structures can be related to unavoidable temperature variations during STM measurements (on the order of 1\%/K), drift and creep artifacts. However, these possible uncertainty contributions can be completely excluded by determining the periodicity of stepped arrays and considering the period-dependent suppression of the Fourier peak intensities, since this approach works even without additional image correction.

\section{Conclusion}

We have demonstrated that the periodicity of the regular, atomically precise triple-step arrays on stepped Si($h\,h\,m$) wafers can be determined with precision approaching the interatomic distance by analyzing the fast Fourier transform (FFT) patterns of differential maps obtained from raw STM topography images using the difference-of-Gaussians (DoG) approach. We have predicted the suppression of specific Fourier peaks at nonzero $k^{\,}_x$ and $k^{\,}_y = 0$ depending on the period of the repeating stripes with the Si(1\,1\,1)$7\times 7$ reconstruction. This effect arises from commensurability between different periodicities in the $k$-space ($k^{\,}_0/7$ controlled by the local $7\times 7$ reconstructions and $k^{\,}_0/L$ controlled by the period $L$ of the $7\times 7$ stripes in the $x$-direction, where $k^{\,}_0$ is the position of the first-order Fourier peak for the $1\times 1$ lattice). This finding allows us to determine the period of triple-step arrays on stepped surfaces by examining the sequential number(s) of the suppressed Fourier peak(s) in the differential map obtained from a raw topography image using the DoG approach. The method seems rather robust and can be applied directly to distorted and tilted images. Considering the period-dependent suppression of the Fourier peaks, we have argued that the local periodicity of the studied Si($h\,h\,m$) surfaces, which consist of flat Si(1\,1\,1) terraces of one $7\times 7$ unit cell and an array of triple steps, equals 5.99 nm and thus corresponds to the Si(8\,8\,11) surface. Such a type of vicinal surface has not been previously considered in the literature.

We should emphasize that the described approach cannot be directly applied to topography images due to (i) finite tilt of the flat terraces with atomic features with respect to the scanning plane and (ii) quantized changes in visible height. Both factors make the structure of $k$-space more complicated and less convenient for in-depth analysis.

An open question is how the concept of the period-dependent suppression of the Fourier peaks can be used for interpretation of LEED patterns. Indeed, some LEED studies demonstrate the disappearance of some diffraction spots at $k^{\,}_y=0$ \cite{Henzler-03,Chaika-SurfSci-09}; however, the positions of the suppressed LEED spots are not symmetrical with respect to the origin as would be expected for the FFT analysis of the DoG patterns. One can suggest that the relative intensities of the diffraction maxima for the stepped surfaces may additionally depend on the sample normal tilt relative to the incident electron beam direction.

\section{Acknowledgements}

The authors are grateful to Dr. Nikolay V. Abrosimov from Leibniz Institute for Crystal Growth (IKZ) for providing vicinal silicon single-crystal wafers. This work was funded by the Russian State Contracts for Osipyan Institute of Solid State Physics RAS (sample preparation and measurements) and for Institute for Physics of Microstructures RAS  (programming and data analysis). A. Yu. Aladyshkin thanks Ministry for Science and Education of Russian Federation (project 075-15-2024-632).


%
%


\begin{thebibliography}{99}
\bibitem{Comment}
The angle $\theta$ between two planes, characterized by the Miller indices  \mbox{($h^{\,}_1$, $k^{\,}_1$, $l^{\,}_1$)} and \mbox{($h^{\,}_2$, $k^{\,}_2$, $l^{\,}_2$)}, can be determined from the relationship
\begin{eqnarray}
\label{Eq-angle}
\nonumber
\theta = {\rm arccos}\,\left(\frac{h^{\,}_1h^{\,}_2+k^{\,}_1k^{\,}_2+l^{\,}_1l^{\,}_2}{\sqrt{h^{2}_1+k^{2}_1+l^{2}_1}\,
\sqrt{h^{2}_2+k^{2}_2+l^{2}_2}}\right).
\end{eqnarray}
\end{thebibliography}
\end{document}